\documentclass[aps,tightenlines,showpacs]{revtex4}
\usepackage[dvips]{graphicx}

\newcommand{\kk}{{\bf k}}

\newcommand{\xx}{{\bf x}}
\newcommand{\yy}{{\bf y}}

\newcommand{\BE}{\begin{equation}}
\newcommand{\EE}{\end{equation}}
\newcommand{\BA}{\begin{eqnarray}}
\newcommand{\EA}{\end{eqnarray}}
\newcommand{\Tr}{\mathrm Tr}
\newcommand{\lamb}{\lambda_{B}}
\newcommand{\eb}{e_{B}}
\newcommand{\mb}{m_{B}}
\begin{document}

\title{A general  interpolation scheme for thermal fluctuations
in superconductors}

\author{L.  Marotta}
\author{ M. Camarda}
\author{ G.G.N. Angilella}
\author{F. Siringo}
\affiliation{Dipartimento di Fisica e Astronomia, 
Universit\`a di Catania,\\
INFN Sez. di Catania, Lab. MATIS-INFM and CNISM Sez. di Catania,\\
Via S.Sofia 64, I-95123 Catania, Italy}

\date{\today}
\begin{abstract}
We present a general interpolation theory for the phenomenological
effects of thermal fluctuations in superconductors. 
Fluctuations are described by a simple gauge invariant extension of the gaussian
effective potential  for the Ginzburg-Landau static model.
The approach is shown to be a genuine variational method, and to be stationary
for infinitesimal gauge variations around the Landau gauge.
Correlation
and penetration lengths are shown to depart from the mean field
behaviour in a more or less wide range of temperature below the
critical regime, depending on the class of material considered.
The method is quite general and yields a very good interpolation of the experimental data
for very different materials.
\end{abstract}

\maketitle

\section {Introduction}
In a recent paper\cite{camarda} we have shown that  the Gaussian Effective Potential (GEP) 
can describe the behaviour of superconductors,
thus allowing for 
a comparison with the experimental data. 
The GEP has a long 
history\cite{schiff,rosen,barnes,kuti,chang,weinstein,huang,bardeen,peskin,stevenson} 
and has been  discussed
by several authors as a tool for describing the breaking of symmetry
in the standard model of  electroweak interactions\cite{stevenson,ibanez}, but in that context
no real comparison with experimental data will be achievable
until the detection of the Higgs boson. 
Thus High $T_c$
superconductors represent the best way to test the
reliability of the method.
While the comparison with experimental data  is of special importance
as a test for the GEP variational method itself, the method can be regarded as a general
interpolation scheme for the pre-critical region of superconductors: too close to the critical
point for thermal fluctuations being negligible, but not yet in the critical range where universality
sets in. In this pre-critical range, the GEP provides an interpolation between the mean-field behaviour 
and the critical limit. Besides  standard phenomenological parameters the method relies on one free
parameter which characterizes the width of  the pre-critical range and allows for a very good fit of  the
experimental data for very different materials ranging from cuprates like YBa$_2$Cu$_2$O$_{7-\delta}$ 
to the double band superconductor  MgB$_2$. The free parameter turns out to be a characteristic
energy scale of  the material.

The general phenomenology of superconductivity
can be described by the standard Anderson-Higgs mechanism:
the supercurrent is carried by pairs of charged fermions whose
non-vanishing expectation value breaks the gauge symmetry, and
endows the gauge bosons with a non-zero mass. Thus, the standard 
Ginzburg-Landau (GL) effective Lagrangian  provides the best 
framework for a general description of the phenomenology.
Moreover, as the GL action can be seen as a power expansion of
the exact action around the critical point, the GL action must be
recovered by any microscopic theory at least around the transition.
Therefore, whatever the origin of the microscopic pairing mechanism,
the GL action is a sound starting point for a general interpolation 
scheme of superconductivity. 
Of course we cannot trust the mean-field
approach to the GL effective theory, and we expect that in 
high $T_c$ superconductors many
unconventional properties should have to do with the breaking down of
the simple mean-field picture. 
Actually, unconventional superconductors
are characterized by a very small correlation length $\xi$, which
allows the experimentalists to get closer to the critical point
where the thermal fluctuations cannot be neglected and the 
mean-field approximation is doomed to fail.
As far as we know there is no clear evidence that the critical 
universal behaviour can always be reached in the cuprate superconductors:
a weakly charged superfluid regime has been predicted and observed\cite{fisher,schneider92,schneider93,kamal} in
nearly optimally doped samples, while a charged critical behaviour has been reported for
the underdoped samples\cite{schneider04}. Recently
the superconducting pyrochlores have been shown to open up a window onto the
charged critical regime\cite{schneider05}.
However it is out of doubt that an intermediate range of temperature is
now accessible, where thermal fluctuations are not negligible
even when the sample is still out of the truly critical regime.
Thus, in order to describe some unconventional properties of
high $T_c$ superconductors, we need to incorporate the role
of thermal fluctuations, but unfortunately we cannot rely
on the standard renormalization group methods\cite{herbut} which would only
describe the limiting universal behaviour. 
We need an approximate approach to the GL action
for the non-universal regime where the behaviour depends on the
physical parameters of the sample, and we would prefer a 
non-perturbative approximation in order to deal with any coupling, whatever its strengthness.

As a toy model for electroweak interactions,
the scalar electrodynamics in four dimensions has been studied by
Iba\~nez-Meier et al.\cite{ibanez} who computed the GEP by use of 
Cartesian coordinates for the field components. This choice breaks
the U(1) symmetry of the field and gives rise to an unphysical,
and undesirable, massive degree of freedom. However,
in three space dimensions, the U(1) scalar electrodynamics is equivalent
to the standard static GL effective model of superconductivity, and
the GEP becomes a variational tool for superconductivity.
In this paper, we show that the unphysical degree of freedom can be avoided
by use of polar coordinates for the field components. 
Moreover, the choice of a U(1) symmetric Gaussian functional restores the gauge
symmetry and makes the theory gauge invariant for infinitesimal gauge changes
around the Landau gauge.
In a previous work we discussed the same approximation in unitarity
gauge\cite{camarda}. The method has  been extended to two-dimensional systems
by Abreu et al\cite{abreu} in order to study film superconductors.
Unfortunately, we found out that the claimed exact\cite{kleinert0} integration of the longitudinal component
of the electromagnetic vector field $A_\mu$ is only approximatively exact.
Actually, as we show in the present paper,  that approximation is a variational one and its accuracy can be
checked by standard variational arguments. Indeed, by a direct comparison of the energies, we show that
the present method yields a better result compared with the Cartesian coordinate GEP of 
Iba\~nez-Meier et al\cite{ibanez}.

The variational method
provides a way to evaluate important phenomenological quantities
like the correlation length $\xi$ and the
penetration depth $\lambda$ that emerge as the solution of a set of coupled equations.
The GL parameter $\kappa_{GL}=\lambda/\xi$, whose critical behaviour has been addressed
by Herbut and Tesanovic\cite{herbut}, is here shown to be  described by a very simple 
relation in the pre-critical regime where it
is found to be temperature-dependent, in contrast to the mean-field result.
Thus, the GL parameter has been chosen as a useful measure of thermal fluctuations: its
dependence on temperature is a sign of the breaking down of the mean-field approximation
which predicts a constant $\kappa_{GL}$. Unfortunately, experimental data on the temperature
dependence of  the GL parameter are quite fragmentary in the literature. The behaviour of
$\xi$ and $\lambda$ can be extracted by the knowledge of the critical magnetic fields, but
experimental limits on these data have narrowed our chances of a wide comparison with
experiments. Up to now, our GEP has shown the capability of fitting the GL parameter for
all the materials that we have been able to test.

The paper is organized as follows: in section II the GL action and
partition function are reported and the symmetry of the model is discussed
together with the conditions for the gauge invariance of  the variational
functional; in section III the integration of  the longitudinal vector field is shown
to be a variational approximation; in section IV the GEP is evaluated in polar coordinates
and the result is compared with the Cartesian coordinate method of Ref.\cite{ibanez};
in section V the variational equations are solved for some set of phenomenological
parameters and the GL parameter is compared with  the available experimental data;
some comments and final remarks on the interpolation method  are reported.

\section {The GL action and gauge invariance}

Let us consider the standard static GL action\cite{kleinert}
\BE
S=\int d^{3}x \left[\frac{1}{4} F_{\mu \nu}F^{\mu
\nu}+\frac{1}{2}(D_{\mu}\phi)^{*}(D^{\mu}\phi)+
\frac{1}{2}m^{2}_{B}\phi^{*}\phi+
\lambda_{B}(\phi^{*}\phi)^{2}\right].
\label{gl}
\EE
Here $\phi$ is a complex (charged) scalar field, its covariant
derivative is defined according to
\BE
D_{\mu} = \partial_{\mu}+ie_{B} A_{\mu}
\label{derivative}
\EE
and $\mu,\nu=1,2,3$ run over the three space dimensions.
The components of the magnetic field  
$F_{\mu\nu}=\partial_\mu A_\nu-\partial_\nu A_\mu$ satisfy
\BE
\frac{1}{2}F_{\mu \nu}F^{\mu \nu} =\vert{\nabla}\times
{\mathbf A}\vert^2
\EE
and the partition function is defined by the functional integral
\BE
Z=\int D[\phi,\phi^*,A_\mu]e^{-S}.
\label{z}
\EE
The action $S$ in Eq.(\ref{gl}) has a local $U(1)$ symmetry as it is invariant
for a local gauge transformation 
\BE
{\bf A} \to
{\bf A}+{\bf \nabla} \chi(x)
\EE
\BE
\phi\to\phi e^{\displaystyle{-ie_B \chi(x)}}
\EE
where $\chi(x)$ is an arbitrary function. The integration over $\bf A$ is then redundant in Eq.(\ref{z}) and
a gauge fixing term must be inserted in order to make the partition function finite.
According to the standard De Witt-Faddeev-Popov method\cite{dewitt} the partition function can be
written as
\BE
Z=\int D[\phi,\phi^*,A_\mu]e^{-S}e^{-S_{fix}}.
\label{zfix}
\EE
where the gauge fixing action is
\BE
S_{fix}=\int d^3 x {1\over {2\epsilon}}f^2
\label{sfix}
\EE
and $f(A)=0$ is an arbitrary gauge constraint. $Z$ is invariant for any change of the parameter $\epsilon$ and
of  the constraint $f$. With some abuse of language, this invariance property is referred to as gauge
invariance while it is a more general invariance as $Z$ does not depend on the shape of the weight factor which
has been added in Eq.(\ref{zfix}).
Only for $\epsilon\to 0$ the weight factor $\exp(-S_{fix})$ becomes a $\delta$-function which enforces
the constraint $f=0$ on the vector field ${\bf A}$.  Thus gauge invariance denotes the invariance of
the theory for any change of the constraint $f=0$ in the limit $\epsilon\to 0$. This is a weaker condition,
but unfortunately even this is not fulfilled by some approximations to the partition function.
In some approximation schemes, such as the perturbative method, the action $S$ is split in two parts
$S=S_0+S_1$ and different approximations are considered for the two parts: for instance $S_0$
might need no approximations while $\exp(-S_1)$ could be expanded as a power series of some small parameter.
Even when $S$ is gauge invariant the approximate $Z$ can result to be gauge dependent.
It is easy to show that if both $S_0$ and $S_1$ are gauge invariant then the approximate $Z$ is gauge
invariant in the sense that in the limit $\epsilon\to 0$ any gauge change $f\to f+\delta f$ leaves the approximate
$Z$ unchanged. Thus in order to obtain a gauge invariant treatment we must take care that the exact action $S$
is split in gauge invariant terms.

The GEP is a sort of optimized first order approximation: in Cartesian coordinates\cite{ibanez}
the action is split according to
\BA
S_0 & = &\int d^{3}x 
\bigg[\frac{1}{4} F_{\mu \nu}F^{\mu\nu}+
\frac{1}{2}(\partial_{\mu}\phi_1)(\partial^{\mu}\phi_1)+
\frac{1}{2}(\partial_{\mu}\phi_2)(\partial^{\mu}\phi_2)+ \nonumber\\
& + &\frac{1}{2}\Omega_1^{2}(\phi_1-\varphi_1)^2+
 \frac{1}{2}\Omega_2^{2}(\phi_2-\varphi_2)^2+
\frac{1}{2}\Delta^{2}{\bf A}^2
\bigg]
\label{gl0}
\EA
where $\phi_1+i\phi_2=\phi$ and $S_1=S-S_0$.
Here $\Omega_i$, $\Delta$ and $\varphi_i$  
are arbitrary variational parameters, and the free energy is evaluated up to first
order in the expansion of $S_1$. The result is then optimized by variation of the free parameters.
When the shifts $\varphi_i$ are non vanishing the action $S_0$ is not $U(1)$ invariant, and the approximate
partition function fails to be gauge invariant.
The lack of symmetry is evident from the appearance of  an unphysical massive Goldstone boson 
(in fact, there are two different mass parmeters $\Omega_1$ and $\Omega_2$).
In the next section we show  by a variational argument that in polar coordinates
the action $S_0$ can be taken as
\BE
S_0=\int d^{3}x 
\left[\frac{1}{4} F_{\mu \nu}F^{\mu\nu}+
\frac{1}{2}(\partial_{\mu}\rho)(\partial^{\mu}\rho)+
\frac{1}{2}\Omega^{2}(\rho-\varphi)^2+
\frac{1}{2}\Delta^{2}{\bf A}^2
\right]
\label{gl0polar}
\EE
where the field is decomposed according to $\phi=\rho\exp(i\gamma)$ and $\rho$ is a $U(1)$ invariant
real field. The explicit integration of the phase $\gamma$ makes the action invariant for rotations
in the $\phi_1$,$\phi_2$ plane.
As before  $\Omega$, $\Delta$ and the shift $\varphi$ are variational parameters.
We still have a non-invariant term quadratic in the vector field. However, if we take
\BE
f=\partial_\mu A^\mu
\EE
any infinitesimal gauge change $f\to f+\delta f$ yields
\BE
A_\mu\to  A_\mu+\partial_\mu \chi
\EE
and up to a surface term
\BE
A^\mu A_\mu\to A^\mu A_\mu-2\chi f+{\cal O} (\delta f^2)
\EE
so that for $\epsilon\to 0$ the constraint $f=0$ restores gauge invariance up to first order in the gauge change
around the Landau gauge $\partial_\mu A^\mu=0$.
This discussion motivates in part the choice of  polar coordinates that gives rise to an explicitly rotational
invariant variational functional and no unphysical massive Goldstone boson. Another motivation arises
from the more pragmatic observation that the free energy is lower with the choice of this polar coordinate
variational functional, and since we show that both methods are genuine variational methods a lower free
energy means an higher accuracy of the results.

\section {GL action in polar coordinates}

In the previous section we have given some motivations for
the choice of polar coordinates in Landau gauge.
In three space
dimensions that gauge is the transverse gauge
$f={\bf \nabla}\cdot{\bf A}$.
A simple coordinate change $\phi\to\rho\exp(i\gamma)$
in the partition function Eq.(\ref{zfix})
yields
\BE
Z=\int D[A_\mu, \rho^2] e^{\displaystyle{-\int d^3x {\cal L}}}
\int D[\gamma] e^{\displaystyle{-\int d^3 x{\cal L}_\gamma}}
\label{zpol}
\EE
where ${\cal L}$ is the phase independent lagrangian 
\BE
{\cal L}=\frac{1}{2} ({\bf \nabla}\times{\bf A})^2
+\frac{1}{2}\partial_{\mu}\rho\partial^{\mu}\rho+
\frac{1}{2}m^{2}_{B}\rho^2+
\lambda_{B}\rho^{4}+\frac{1}{2}e^2_B\rho^2{\bf A}^2+
\frac{1}{2\epsilon} ({\bf \nabla}\cdot{\bf A})^2
\label{L}
\EE
and ${\cal L}_{\gamma}$ is the sum of the Lagrangian terms which 
depend on the phase $\gamma$
\BE
{\cal L}_\gamma=
\frac{1}{2}\rho^2\partial_{\mu}\gamma\partial^{\mu}\gamma+
e_B\rho^2\partial _\mu\gamma A^\mu.
\label{Lgamma}
\EE
If the last term of ${\cal L}_\gamma$ were neglected, the phase $\gamma$ could
be integrated exactly (the integral is Gaussian).
It has been claimed\cite{camarda,abreu,kleinert0} that in transverse gauge that term should
vanish exactly: the argument was that in transverse gauge there is no longitudinal component of
the vector field. In fact, the Fourier transform of the gradient $\partial_\mu \gamma$ is proportional
to the wave vector ${\bf k}$ and this is orthogonal to the Fourier transform of  ${\bf A}$ for
the gauge constraint  $f({\bf k})={\bf k}\cdot{\bf A}({\bf k})=0$. However, this holds only for Fourier
transforms. The last term of ${\cal L}_\gamma$ contains a $\rho^2$ factor which is not
constant, and  thus, in this case, the space integral of this term in the action cannot be
replaced by the simple inverse-space integral of the product of the Fourier transforms
(we could do that  if the factor $\rho$ were constant). Thus we cannot neglect that  term and
the exact integration over the phase $\gamma$ would depend on the field ${\bf A}$.
We stress that any gauge change does not solve the problem either, as an undesirable longitudinal
component of the field ${\bf A}$ would take the place of the phase $\gamma$. For instance in Ref.\cite{kleinert05}
the phase $\gamma$ is gauged away but the quadratic term $e_B^2\rho^2{\bf A}^2$  contains  an implicit
coupling term $e_B^2\rho^2 {\bf  A}_L\cdot{\bf A}_T$ between the transversal component ${\bf A}_T$
and the longitudinal component ${\bf A}_L$ of the vector field. Again the integral of this coupling term
would vanish if $\rho$ were constant (as it is obvious by the same Fourier transform argument discussed above),
but it is not vanishing in general and prevents from an exact integration of the longitudinal component.

We can get rid of the phase by a variational argument: it turns out that the neglection of the
phase terms is not an exact integration, but a good variational approximation.
Let us denote by ${\cal L}_0$ the phase dependent Lagrangian without the last term
\BE
{\cal L}_0={\cal L}_\gamma(e_B=0)=
\frac{1}{2}\rho^2\partial_{\mu}\gamma\partial^{\mu}\gamma.
\label{L0}
\EE
We observe that up to constant factors the exact integration over $\gamma$ yields
\BE
\int D[\gamma] e^{\displaystyle{-\int d^3 x{\cal L}_0}}\sim\prod_x \frac{1}{\rho}.
\label{factor}
\EE
Thus we may write the $D[\rho^2]$ integral in Eq.(\ref{zpol}) as
\BE
Z\sim\int D[A_\mu, \rho] e^{{-\int d^3x {\cal L}}}\>
\left\{{{\int D[\gamma] e^{{-\int d^3 x{\cal L}_\gamma}}}
\over
{\int D[\gamma] e^{{-\int d^3 x{\cal L}_0}}}}\right\}.
\label{zav}
\EE
We define the average over the phase as
\BE
\langle(...)\rangle_\gamma={{\int D[\gamma] e^{-\int d^3 x{\cal L}_0}(...)}\over   
{\int D[\gamma] e^{-\int d^3 x{\cal L}_0}}}   
\label{average}.
\EE
With this notation the exact partition function Eq.(\ref{zpol}) reads
\BE
Z=\int D[A_\mu, \rho] e^{\displaystyle{-\int d^3x {\cal L}}}
\left\langle e^{\displaystyle{-\int d^3 x e_B\rho^2\partial_\mu \gamma A^\mu}}\right\rangle_\gamma
\label{zav2}
\EE
and the convexity of the exponential function ensures that
\BE
Z\geq \int D[A_\mu, \rho] \>e^{\displaystyle{-\int d^3x {\cal L}}}
\> e^{\displaystyle{-\int \left\langle e_B\rho^2 \partial _\mu\gamma 
A ^\mu\right\rangle_\gamma d^3 x}}.
\label{diseq}
\EE
The average in the right hand side vanishes (it is linear in $\gamma$), and
the approximate partition function $Z_p$ 
\BE
Z_p=\int D[A_\mu, \rho] e^{\displaystyle{-\int d^3x {\cal L}}}
\label{zappr}
\EE
satisfies the variational constraint
\BE
Z\geq Z_p
\EE
so that the approximate free energy ${\cal F}_p=-\ln Z_p$ is bounded by the exact free
energy  ${\cal F}=-\ln Z$ 
\BE
{\cal F}_p\geq {\cal F}.
\EE
This bound ensures that, even neglecting the last term of the phase dependent lagrangian
${\cal L}_\gamma$ in Eq.(\ref{Lgamma}), the resulting approximate partition function $Z_p$ 
still gives a genuine variational approximation.

\section {The GEP method in polar coordinates}

In this section we study the GEP method for the polar coordinate partition function
Eq.(\ref{zappr}) with the lagrangian
${\cal L}$ defined according to Eq.(\ref{L}). 
The GEP may be evaluated
by the same $\delta$ expansion method discussed in Ref.\cite{ibanez} and
\cite{stancu} and also reported by Camarda et al.\cite{camarda}. 
Here the GEP represents a variational
estimate of the free energy ${\cal F}_p$.

Inserting a source term for the real field $\rho$ the partition function $Z_p$ reads
\BE
Z_p[j]=\int D[A_\mu, \rho] \>e^{\displaystyle{-\int d^3x {\cal L}}}
\>e^{\displaystyle{-\int d^3x j\rho}}
\label{zj}
\EE
and the free energy  is given by the Legendre transform
\BE
{\cal F}[\varphi]=-\ln Z_p+\int d^3 x j\varphi
\label{F}
\EE
where $\varphi$ is the average value of $\rho$.
As usual we introduce a shifted field
\BE
\tilde{\phi}=\rho-\varphi
\EE
and then we split the lagrangian into two parts
\BE
{\cal L}={\cal L}_0+{\cal L}_{int}
\EE
where ${\cal L}_0$ is the sum of two free-field terms:
a vector field $A_\mu$ with mass $\Delta$ and
a real scalar field $\tilde \phi$ with mass $\Omega$:
\BE
{\cal L}_{0}=\left[
+\frac{1}{2}({\nabla} \times {\bf A})^{2} 
+\frac{1}{2} \Delta^{2} A_\mu A^\mu
+\frac{({\nabla}\cdot{\bf A})^{2}}{2\epsilon}
\right]
+\left[
\frac{1}{2}({\nabla}\tilde{\phi})^{2}
+\frac{1}{2}\Omega^{2}\tilde{\phi}^{2}
\right]
\EE
The interaction then reads
\BA
{\cal L}_{int}&=&
v_{0}+v_{1}\tilde{\phi}+v_{2}\tilde{\phi}^{2}+v_{3}\tilde{\phi}^{3}
+v_{4}\tilde{\phi}^{4}+\nonumber\\
&&+\frac{1}{2}\left(e_{B}^{2}\varphi^{2}-\Delta^{2}\right)A_\mu A^\mu
+e_{B}^{2}\varphi A_\mu A^\mu\tilde{\phi}
+\frac{1}{2}e^{2}_{B}A_\mu A^\mu\tilde{\phi}^{2}
\EA
where
\BA
v_{0} &=& \frac{1}{2}m_{B}^{2}\varphi^{2}+\lambda_{B}\varphi^{4}\\
v_{1} &=& m_{B}^{2}\varphi+4\lambda_{B}\varphi^{3}\\
v_{2} &=& \frac{1}{2}m_{B}^{2}+6\lambda_{B}\varphi^{2}-\frac{1}{2}\Omega^{2}\\
v_{3} &=& 4\lambda_{B}\varphi\\
v_{4} &=& \lambda_{B}
\EA

The first order  expansion of the free energy is
\BA
{\cal F}[\varphi]&=&
\frac{1}{2}\Tr \ln\left[ g^{-1}(x,y)\right]
+\frac{1}{2}\Tr \ln\left[ G_{\mu\nu}^{-1}(x,y)\right]+\nonumber\\
&+&\int d^3 x\left\{
v_0+v_2 g(x,x)+3v_4 g(x,x)^2+\frac{1}{2} e_B^2
\left(g(x,x)+\varphi^2-\Delta^2\right)G_{\mu\mu}(x,x)\right\}
\EA
where $g(x,y)$ is the free-particle propagator for the scalar field,
and $G_{\mu\nu}(x,y)$ is the free-particle propagator for the
vector field 
\BE
G_{\mu\nu}^{-1}(x,y)=\int \frac{d^3 k}{(2\pi)^3}
e^{-i\kk(\xx-\yy)}\left[\delta_{\mu\nu}(k^2+\Delta^2)+
\left(\frac{1}{\epsilon}-1\right) k_\mu k_\nu\right].
\EE
In the limit $\epsilon\to 0$, up to an additive constant
\BE
\Tr \ln\left[G_{\mu\nu}^{-1}(x,y)\right]=2{\cal V}
\int \frac{d^3 k}{(2\pi)^3} \ln (k^2+\Delta^2)
\EE
where $\cal V$ is the total volume.
Dropping all constant terms, the free energy density
$V_{eff}={\cal F}/{\cal V}$ (effective potential) reads
\BA
V_{eff}[\varphi] &=& I_{1}(\Omega)+2I_{1}(\Delta)+\nonumber\\
&&+ \left[
\lambda_{B}\varphi^{4}+\frac{1}{2}m_{B}^{2}\varphi^{2}+\frac{1}{2}
\left\{
m^{2}_{B}-\Omega^{2}+12\lambda_{B}\varphi^{2}+6\lambda_{B}I_{0}(\Omega)
\right\} I_{0}(\Omega) \right] \nonumber\\
&&+\left( e_B^{2}\varphi^{2}+e_B^{2}I_{0}(\Omega)-\Delta^{2} \right)
I_{0}(\Delta)
\label{veff}
\EA
where the divergent integrals $I_n$ are defined according to
\BE
I_0 (M)=\int\frac{d^3 k}{(2\pi)^3} \frac{1}{M^2+k^2}
\EE
\BE
I_1 (M)=\frac{1}{2}\int\frac{d^3 k}{(2\pi)^3} \ln(M^2+k^2)
\EE
and are regularized by insertion of a finite cut-off $k<\Lambda$.

The free energy (\ref{veff}) now depends on the mass parameters
$\Omega$, $\Delta$ and on the field shift $\varphi$. These are the variational
parameters that must be determined by the minimization of the
energy density $V_{eff}$. At the stationary point $V_{eff}$ is the
GEP and the mass parameters  give the inverse correlation lengths
for the fields, the so called coherence length $\xi=1/\Omega$ and 
penetration depth $\lambda=1/\Delta$. The field shift $\varphi$ is the
order parameter of the phase transition: when $\varphi\not=0$ at the
minimum of  $V_{eff}$ the $U(1)$ symmetry is broken in the ground
state and the system is in the superconducting phase.

The stationary conditions
\BE
\frac{\partial V_{eff}}{\partial\Omega^{2}}=0
\EE
\BE
\frac{\partial V_{eff}}{\partial\Delta^{2}}=0
\EE
give two coupled gap equations:
\BE
{\Omega}^{2}=12 \lambda_{B}
I_{0}({\Omega})+m^{2}_{B}+12 \lambda_{B}\varphi^{2}+2
e^{2}_{B}I_{0}({\Delta})
\label{gap1}
\EE
\BE
{\Delta}^{2}=e^{2}_{B}\varphi^{2}+e^{2}_{B}I_{0}({\Omega}).
\label{gap2}
\EE
For any $\varphi$ value the equations must be solved 
numerically, and the minimum point values $\Omega$
and $\Delta$ must be inserted back into Eq.(\ref{veff})
in order to get the Gaussian free energy 
$V_{eff}(\varphi)$
as a function of the order parameter 
$\varphi$.
For a negative and
small enough $m_B^2$, we find that $V_{eff}$ has a minimum at a 
non zero value of $\varphi=\varphi_{min}>0$, thus indicating that the
system is in the broken-symmetry superconducting phase.
Of course the masses $\Omega$, $\Delta$ only take their physical
value at the minimum point of the free energy $\varphi_{min}$.
This point is found  by requiring that
\BE
\frac{\partial V_{eff}}{\partial\varphi^{2}}=0
\label{phimin}
\EE
where as usual the partial derivative is allowed as far as the
gap equations (\ref{gap1}),(\ref{gap2}) are satisfied\cite{stevenson}.
The condition (\ref{phimin}) combined with the gap equation (\ref{gap1})
yields the very simple result
\BE
\varphi_{min}^{2}=\frac{{\Omega^{2}}}{8\lambda_{B}}.
\label{phimin2}
\EE
However, we notice that here the mass $\Omega$ must be found
by solution of the coupled gap equations. Thus Eqs.(\ref{phimin2}),
(\ref{gap1}) and (\ref{gap2}) are a set of
coupled equations and must be solved together in order to find
the physical values for the correlation lengths and the order
parameter.

Insertion of Eq.(\ref{phimin2}) into Eq.(\ref{gap2}) yields the
simple relation for the GL parameter $\kappa_{GL}$
\BE
\kappa_{GL}^{2} = \left( \frac{\lambda}{\xi}
\right)^{2}=\kappa_0 \>{1 \over
\displaystyle 1+\frac{I_{0}(\Omega)}{\varphi_{min}^{2}} }
\label{kappa}
\EE
where $\kappa_0=e_B^2/(8\lambda_B)$ is the mean-field GL parameter
which does not depend on temperature. As discussed by Camarda et al.\cite{camarda},
Eq.(\ref{kappa}) shows that
the GL parameter is predicted to be temperature dependent
through the non trivial dependence of $\Omega$ and $\varphi_{min}$.
At low temperature, where the order parameter $\varphi_{min}$ is
large, the deviation from the mean-field value $\kappa_0$ is
negligible. Conversely, close to the critical point, where the
order parameter is vanishing, the correction factor in Eq.(\ref{kappa})
becomes very important. The deviation from the mean-field prediction
$\kappa_{GL}=\kappa_0$ also depends on the cut-off parameter
$\Lambda$ as for $\Lambda\to 0$ all the effects of  fluctuations 
vanish together with the integrals $I_n$. This parameter defines 
another length scale $\ell=1/\Lambda$, which is a characteristic of
the material.

A test of the present method comes from the direct comparison with
the cartesian coordinate GEP discussed by Iba\~nez-Meier et al.\cite{ibanez}.
Both methods are genuine variational methods and a direct comparison of
the free energy density gives information on the accuracy. In Fig.1 we report
our polar coordinate GEP together with the three-dimensional version of the
cartesian coordinate GEP. The polar coordinate GEP yields a lower free energy
density thus indicating that, besides being more symmetric and  appealing (there
are no unphysical degree of freedom and no massive gauge bosons), the method
is also more effective.

\section {Interpolation of the experimental data}
As we mentioned earlier, another appealing aspect of the GEP approximation 
in three space-time dimensions resides in the chance to compare its 
results with the phenomenology of high $T_c$ 
superconductors. For this comparison to be consistent we need to fix the bare 
parameters of the GL action; following \cite{camarda}, we can use the standard
derivation of the GL action (\ref{gl})(see \cite{kleinert}) to find 
a connection between microscopic {\it first-principle} quantities and phenomenologic
bare parameters. Thus, we have:
\BE
e_{B}=\frac{2e}{\hbar c}\sqrt{k_{B}T_{c}\xi_{0}}
\EE
where $2e$ represents the charge of a Cooper pair, $T_c$ is the critical temperature 
and $\xi_0$ the zero temperature coherence length.

The knowledge of the zero temperature coherence length and penetration depth 
enables us to fix the other parameters; regarding the bare mass parameter $m_B^2$
 as a linear function of temperature, we obtain:
\BE
m_B^{2}=m_{c}^{2}+\left[1-\frac{T}{T_{C}}\right](m_{0}^{2}-m_{c}^{2})
\EE
where $m_0^2$ is the value which is required in order to find 
$\Omega=1/\xi_0$ from the gap equation (\ref{gap1}) at $\varphi=\varphi_{min}$,
and $m_c^2$ is the value of $m_B^2$ at the transition point.
The mean-field approximation predicts that  $m_c^2=0$, but the effect of the fluctuations 
is to shift the transition point to a negative non-vanishing  $m_B^2$ value. 

Finally, the bare coupling $\lambda_B$ is
fixed through Eq.(\ref{gap2}) by requiring that at zero temperature 
(i.e. for $m_B^2=m_0^2$)
the penetration depth $\lambda_0=1/\Delta$.\\
The cutoff $\Lambda$ still remains to be fixed; thus, the interpolation scheme 
for the superconducting properties we have to deal with contains one 
free parameter (i.e. $\Lambda$), which is a characteristic energy scale of the sample, and 
will be determined by a direct fit of the experimental data.

We have compared our theoretical results with the experimental GL parameter for three 
different materials: Tl$_{2}$Ca$_{2}$Ba$_{2}$Cu$_{3}$O$_{10}$ ($T_c = 121.5$~K, $\xi_0=1.36$ nm, 
$\kappa_0=100$) \cite{experiment}, 
YBa$_2$Cu$_2$O$_{7-\delta}$ ($T_{c}=62.2$ K, $\xi_{0}= 1.65$nm, $\kappa_{0}=187$) \cite{vandervoort} 
and 
MgB$_2$ ($T_{c}=38.6$K, $\xi_{0}= 3.9$nm, $\kappa_{0}= 31.66$) \cite{xu}. For the last two 
materials we have obtained the GL parameter by the relation \cite{poole}:
\BE
\frac{H_{c2}}{2H_{c1}}=\frac{\kappa^{2}}{\ln\kappa}
\EE
(where $H_{c1}$ and $H_{c2}$ represent the lower and upper critical fields,respectively),
and extrapolated the zero temperature  phenomenological values needed to fix the bare parameters by a linear fit.

The small number of available experimental data is due to the difficulty of performing 
measurements of the coherence length and penetration depth (or, alternatively, 
of the two critical fields) up to the pre-critical region.

In Fig.2, Fig.3 and Fig.4 the comparisons between our interpolation curves, as obtained  by Eq.(\ref{kappa}),
and the data relative to the three materials mentioned above, are shown.\\
It can be noted that the experimental GL parameter is almost constant at low temperatures, 
according to the mean-field description, while it shows a sharp temperature-dependence 
(at $T/T_{c}\approx 0.8$ for Tl$_{2}$Ca$_{2}$Ba$_{2}$Cu$_{3}$O$_{10}$, at $T/T_{c}\approx 0.9$ 
for YBa$_{2}$Cu$_{2}$O$_{7_-\delta}$,
at $T/T_{c}\approx 0.6$ for  MgB$_{2}$)  in the pre-critical region: it is in this region 
that the mean-field approximation breaks down, while our model shows its effectiveness.
We would like to emphasize that the GEP approximation is able to reproduce the GL parameter 
behaviour for classes of compounds (cuprates and diborides) with very different microscopic 
structures and, probably, different \emph{pairing mechanism}. These differences are reflected 
in the cutoff values we used to fit the data: they have the same order of magnitude 
($\Lambda\xi_0=20$ for Tl$_{2}$Ca$_{2}$Ba$_{2}$Cu$_{3}$O$_{10}$, 
$\Lambda\xi_0=5$ for YBa$_{2}$Cu$_{2}$O$_{7_-\delta}$)
 for the two cuprates, while the cutoff used for the diboride ($\Lambda\xi_0=900$ for MgB$_{2}$) is about 
two order of magnitude greater. In this sense, the parameter $\Lambda$ seems to be a characteristic 
energy scale of the sample, 
related to its microscopic structure, and, as a consequence, to the size of the region where fluctuations cannot be 
neglected.

Finally, we want to remark that, very close to the critical point, 
some universal
behaviour should be expected and the role of thermal fluctuations
becomes too important to be dealt with by the present method.
In fact the GEP always predicts a weak first order transition at the critical
point even for the neutral superfluid (real scalar theory)\cite{siringo}.
This is not a problem for the interpolation of the experimatal data as the
difference only arises in a very narrow range of temperature at the transition point.
Thus the method seems to be suitable for interpolating 
the pre-critical region where the order of the transition does not make any difference.
Any extension to the critical range would require the proper inclusion of vortex fluctuations
which are expected to play a major role at criticality\cite{camarda2}, and have been recently shown to change
the order of the transition in the GL model\cite{kleinert05}. 
We also mention that a variational perturbative extension of the gaussian
approximation has been shown to be able to reach the critical regime
\cite{kleinert2}

In summary we have shown that  
a polar decomposition of the complex field in the action has the merit of
allowing the integration of the longitudinal vector field (by a further variational 
approximation), is better than the Cartesian decomposition in \cite{ibanez}
and, at variance with the mean-field approximation, yields a  one-parameter interpolation scheme which
fits very well all the available data, in the whole range of the accessible temperatures.
Moreover, the variational character of the model, allowing the study of any kind 
of coupling, makes the GEP a suitable tool for the inspection of other 
physical systems, such as superfluids. 

As the GEP provides a nice way to interpolate the experimental
data beyond the mean-field regime, we expect the method to
be reliable for the description of symmetry breaking in $3+1$
dimensions where the scalar electrodynamics may be regarded
as a toy model for the standard electro-weak theory.

\begin{acknowledgments}
We are grateful to Prof. H. Kleinert for valuable discussions and comments
on the present paper.
\end{acknowledgments}

\begin{figure}[ht]
\includegraphics[height=15cm, width=12cm]{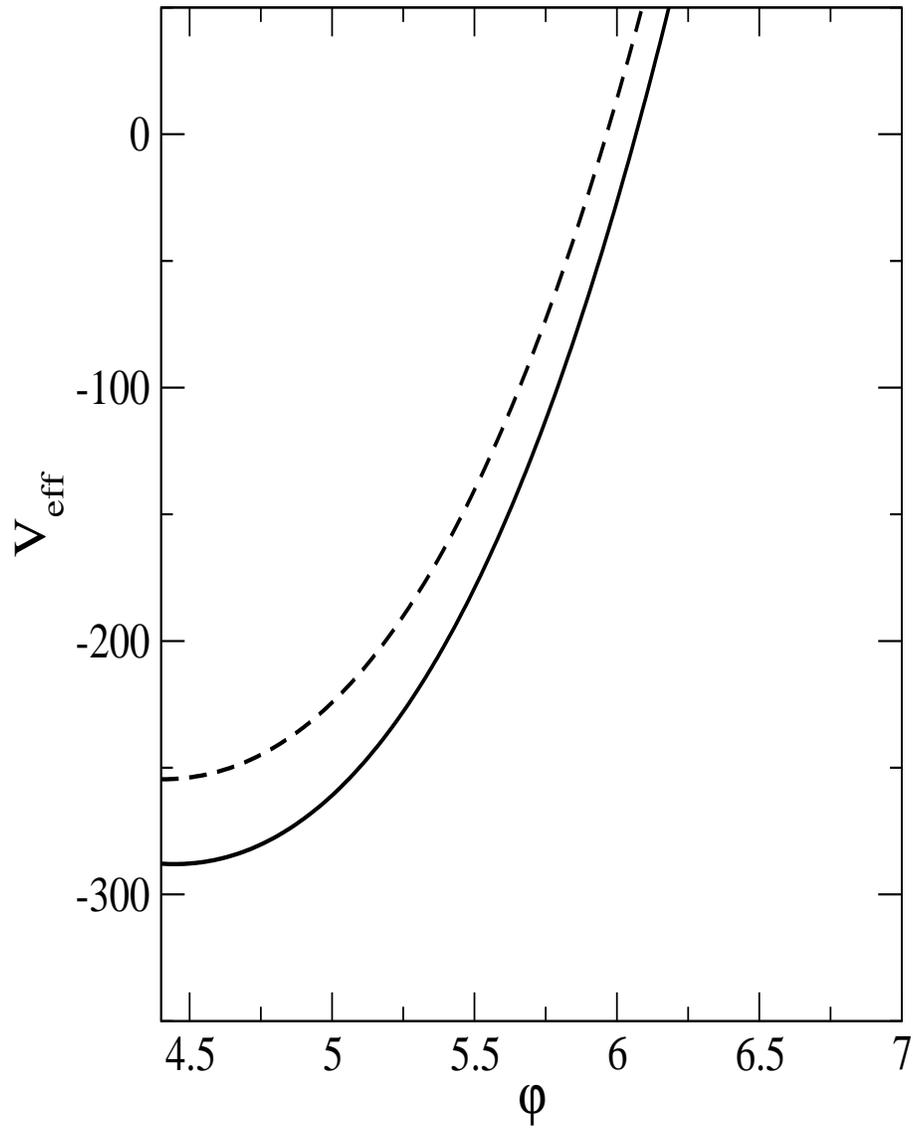}
\caption{\label{Fig1}
Comparison between the {\it polar} GEP (solid line) and the {\it Cartesian} GEP {dashed line}. The value of the 
parameters are: $\mb^{2}/ \lamb^{2}=-80$, $\Lambda/\lamb=10$, $\eb/\lamb=0.001$. Note that the origin 
of the $\varphi$ axis coincides with the minimum of the {\it Cartesian} 
GEP to emphasize that our polar decomposition of the field 
leads to a lower effective potential.}
\end{figure}

\begin{figure}[ht]
\includegraphics{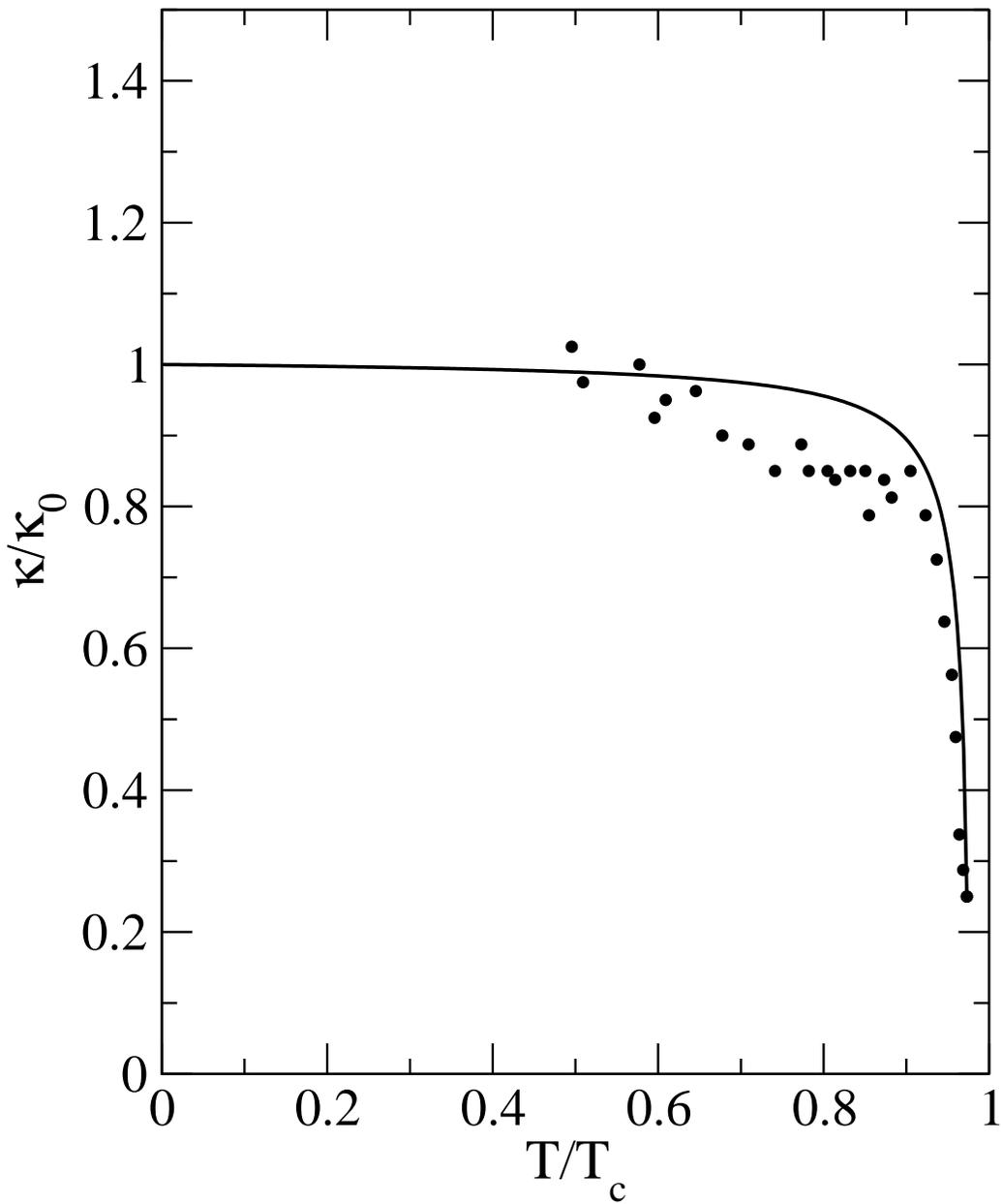}
\caption{\label{Fig2}
The GL parameter according to Eq.(\ref{kappa}) for $\kappa_0=100$,
$\xi_0=1.36$ nm, $T_c=121.5$ K and $\Lambda\xi_0=20$ (full line). 
The circles are the experimental data of Ref.\cite{experiment}  for 
Tl$_{2}$Ca$_{2}$Ba$_{2}$Cu$_{3}$O$_{10}$.}
\end{figure}

\begin{figure}[ht]
\includegraphics{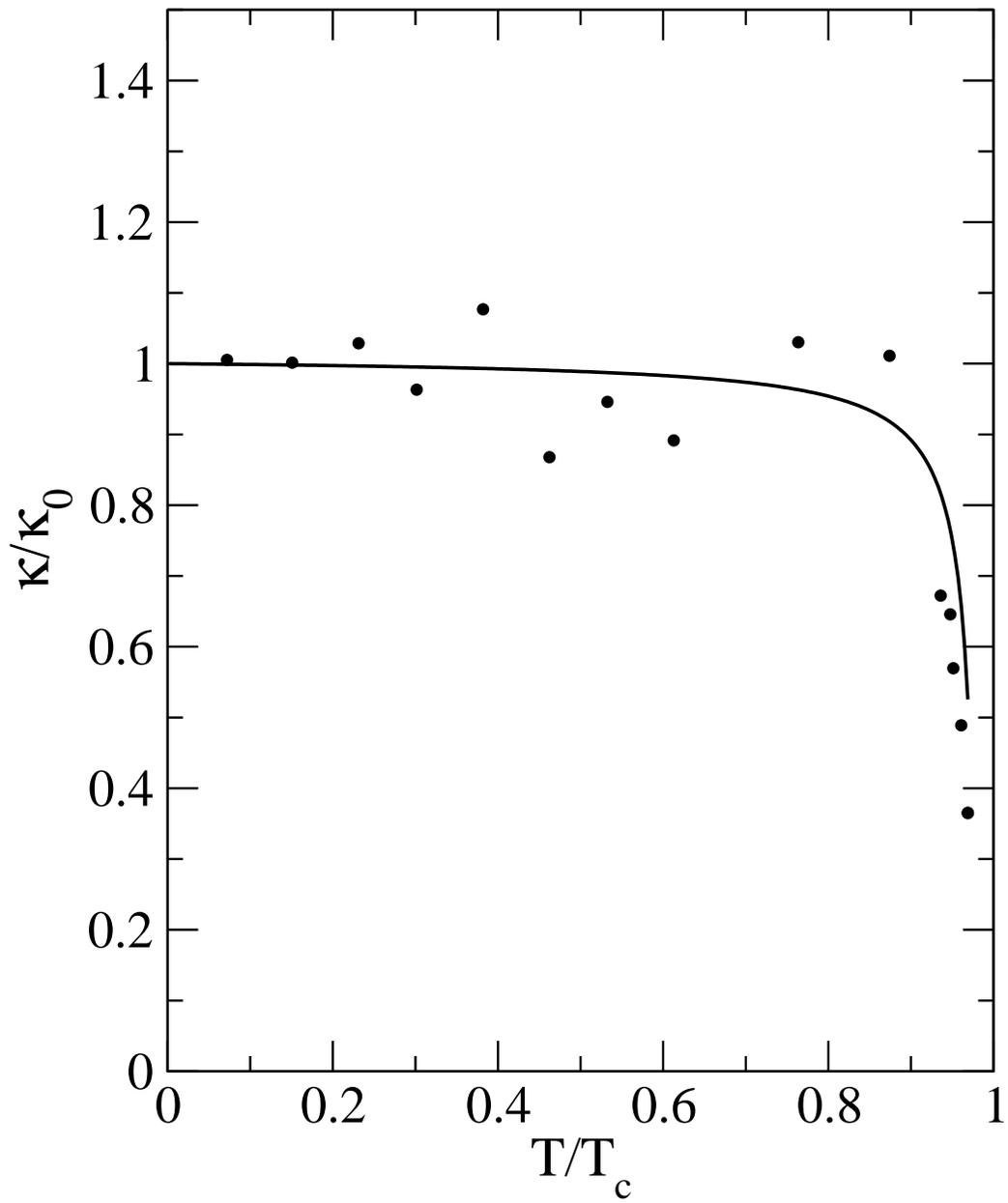}
\caption{\label{Fig3}
The GL parameter according to Eq.(\ref{kappa}) for $\kappa_0=187$,
$\xi_0=1.65$ nm, $T_c=62$ K and $\Lambda\xi_0=5$ (full line). 
The circles are the experimental data of Ref.\cite{vandervoort} for 
YBa$_{2}$Cu$_{2}$O$_{7_-\delta}$.}
\end{figure}

\begin{figure}[ht]
\includegraphics{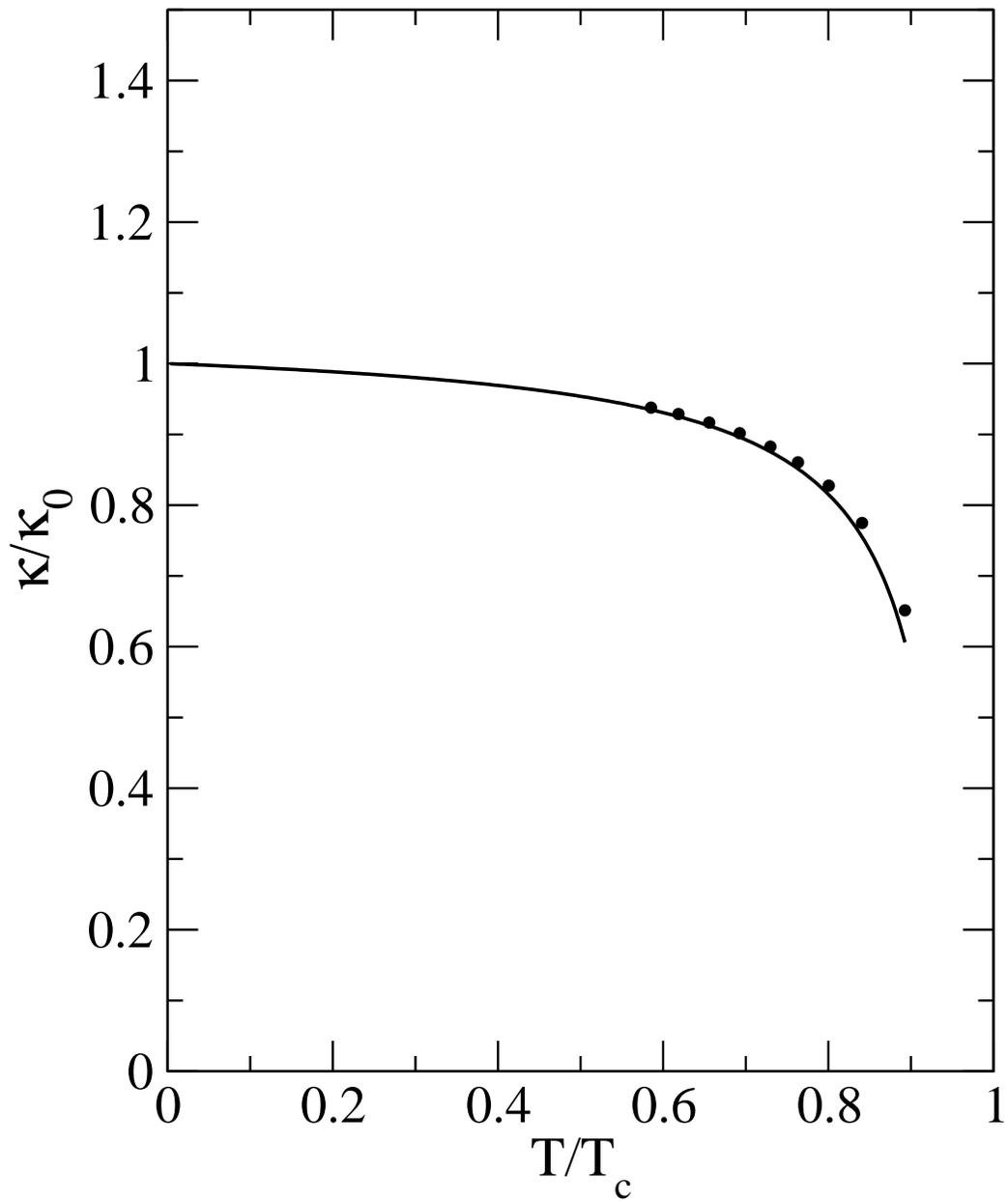}
\caption{\label{Fig4}
The GL parameter according to Eq.(\ref{kappa}) for $\kappa_0=31.66$,
$\xi_0=3.9$ nm, $T_c=38.6$ K and $\Lambda\xi_0=900$ (full line). 
The circles are the experimental data of Ref.\cite{xu}  for  MgB$_2$.}
\end{figure}

\end{document}